\documentclass[sigconf,nonacm]{acmart}

\def\BibTeX{{\rm B\kern-.05em{\sc i\kern-.025em b}\kern-.08emT\kern-.1667em\lower.7ex\hbox{E}\kern-.125emX}}

\usepackage{amsmath,amssymb,amsfonts}
\usepackage{algorithmic}
\usepackage{graphicx}
\usepackage{textcomp}
\usepackage{xcolor}
\usepackage{color,colortbl}
\definecolor{Gray}{gray}{0.9}
\usepackage{todonotes,graphicx,tabu,tabularx}
\usepackage{xspace}
\usepackage{bm}
\usepackage{color}
\usepackage{breakurl}
\usepackage[labelformat=simple]{subcaption}

\usepackage{natbib}

\usepackage{balance}
\renewcommand\footnotetextcopyrightpermission[1]{} 
\begin{document}

\graphicspath{ {images/} }
\newcommand{\sysname}{ModiPick\xspace}

\title{\sysname: SLA-aware Accuracy Optimization For Mobile Deep Inference}

\author{Samuel S. Ogden}
\email{ssogden@wpi.edu}
\affiliation{
  \institution{Worcester Polytechnic Institute}
  \streetaddress{100 Institute Ave.}
  \city{Worcester}
  \state{Massachusetts}
  \postcode{01609}
 }
 
\author{Tian Guo}
\email{tian@wpi.edu}
\affiliation{
  \institution{Worcester Polytechnic Institute}
  \streetaddress{100 Institute Ave.}
  \city{Worcester}
  \state{Massachusetts}
  \postcode{01609}
}

\renewcommand{\shortauthors}{Ogden, et al.}

\begin{abstract}

Mobile applications are increasingly leveraging complex deep learning models to deliver features, e.g., image recognition, that require high prediction accuracy. 
Such models can be both computation and memory-intensive, even for newer mobile devices, and are therefore commonly hosted in powerful remote servers. 
However, current cloud-based inference services employ static model selection approach that can be suboptimal for satisfying application SLAs (service level agreements), as they fail to account for inherent dynamic mobile environment.

We introduce a cloud-based technique called \sysname that dynamically selects the most appropriate model for each inference request, and adapts its selection to match different SLAs and execution time budgets that are caused by variable mobile environments. The key idea of \sysname is to make inference speed and accuracy trade-offs at runtime with a pool of managed deep learning models. 
As such, \sysname masks unpredictable inference time budgets and therefore meets SLA targets, while improving accuracy within mobile network constraints. We evaluate \sysname through experiments based on prototype systems and through simulations. We show that \sysname  achieves comparable inference accuracy to a greedy approach while improving SLA adherence by up to 88.5\%.

\end{abstract}

\keywords{Mobile application, DNN inference service, DNN model management, performance optimization}

\maketitle

\section{Introduction} 
\label{sec:intro}

\begin{figure*}[t]
    \centering
    \includegraphics[width=\textwidth]{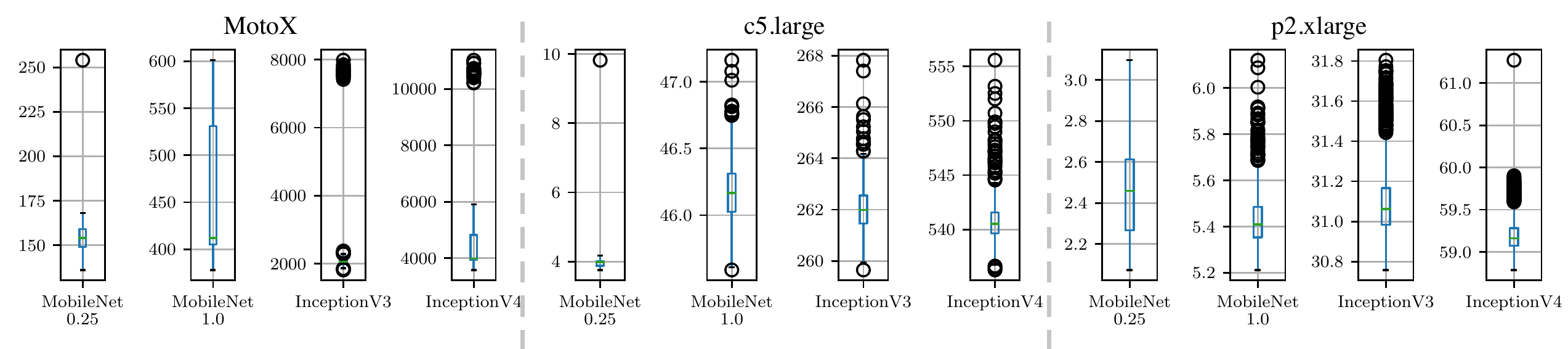}
    \caption{Comparison of inference latency of on-device and cloud-based inference with four state-of-the-art CNN models. \textnormal{Although mobile-optimized model \texttt{MobileNet 0.25} only takes an average of 150ms to run on a mobile phone, it takes up to 26X longer to run a twice accurate \texttt{InceptionV4} model. In addition, the \texttt{InceptionV4} runs in just over 59ms on the \texttt{p2.xlarge} GPU-accelerated server, over 2.5X faster than the \texttt{MobileNet 0.25} on the MotoX. Our measurement identifies the need for cloud-based inference for mobile applications that benefit from highly accurate inference results and demonstrates the potential of cloud-based inference for enabling inference latency and accuracy trade-offs.}}
    \label{bg:modelruntimes}
\end{figure*}

Today mobile applications are increasingly powered by deep learning models, providing rich features such as real-time language translation, image recognition and personal assistants~\cite{NIPS2013_5004, DBLP:journals/corr/WuSCLNMKCGMKSJL16, siri.deeplearning}. 
Unlike traditional mobile applications that rely on much simpler models~\cite{8454307, Ravi:2005:ARA:1620092.1620107}, those new features often require access to ``deeper" models~\cite{resnet, nasnet} that can take unreasonable amount of execution time on mobile hardware~\cite{DBLP:journals/corr/Guo17a, MODI, tensorflow_models} (see Figure~\ref{bg:modelruntimes}). 
Currently, to use deep inference, mobile applications can either resort to model optimization techniques~\cite{DBLP:journals/corr/IandolaMAHDK16, DBLP:journals/corr/ZophVSL17, DBLP:journals/corr/RastegariORF16, DBLP:journals/corr/HowardZCKWWAA17, caffe2, google_tensorflow:lite, Jouppi:2017:IPA:3079856.3080246, pixel2.tpu} that often sacrifice accuracy for improved on-device inference time or leverage cloud-based inference services~\cite{DBLP:journals/corr/ChenLLLWWXXZZ15, tensorflow_serving, clipper}. 
As the need to support more complex application scenarios emerge, various cloud-based serving platforms have become a preferable option for achieving high inference accuracy.

\begin{figure}[t]
\centering
\includegraphics[width=\columnwidth]{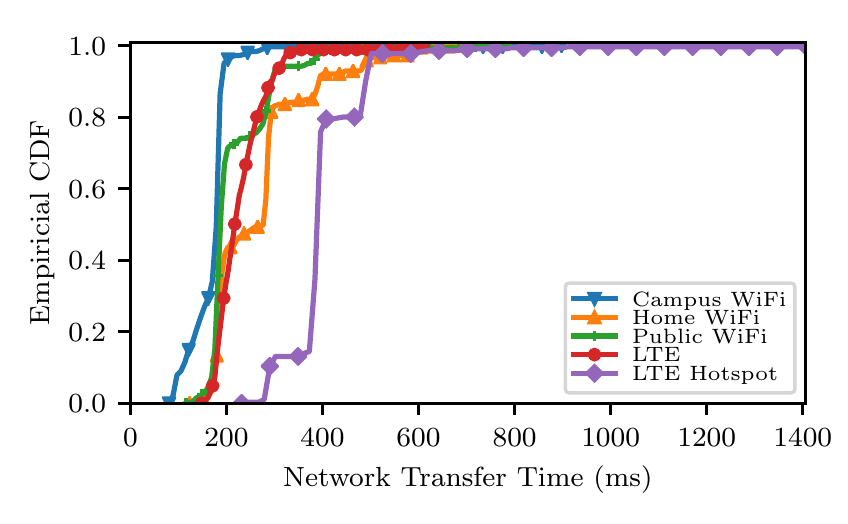}
\caption{Impact of dynamic mobile network condition on end-to-end mobile inference latency. 
\textnormal{We empirically measured the cloud-based mobile deep inference time using our Android-based benchmark application. The network transfer time shown here is obtained by subtracting on-server inference execution from the end-to-end inference response time. Our results reveal two key insights: (1) different network types show significantly different degrees of network latency; (2) even using the same network type, network latency can vary significantly.}}
\label{fig:background:in-cloud}
\end{figure}

Although these proposed general platforms address some fundamental issues in the model post-training phase, they are less effective in serving mobile inference requests. The crux of this inefficiency stems from not considering the challenges that are unique to supporting deep inference requests for mobile applications. \emph{First}, mobile devices are in a much more dynamic network environment than traditional devices, ranging from high-latency cellular connections to high-speed WiFi connections. As shown in Figure~\ref{fig:background:in-cloud}, the median network latency when using LTE hotspot can be twice as much as using campus WiFi. Therefore, mobile devices can take varying amount of time to transfer input data needed for inference tasks. \emph{Second}, these deep learning mobile applications are often user facing and thus have stricter SLA requirements.
User-facing foreground applications, for example generating labels for images or translating captions, have more stringent time requirements than applications running in the background, such as re-processing images in gallery application. 
Consequently, it is important to be aware of and distinguish latency-sensitive mobile inference requests. 
\emph{Third}, even for the same type of inference requests, mobile devices tend to request inference on raw data of different resolutions due to inherently different mobile sensor capacities, such as different image sensors~\cite{iphoneX,nexus5}. Therefore, input data preprocessing can take vastly different amounts of time which can be further exacerbated by heterogeneous mobile processing capacity~\cite{DBLP:journals/corr/Guo17a}. 

\begin{figure}[t]
\centering
\includegraphics[width=0.9\columnwidth]{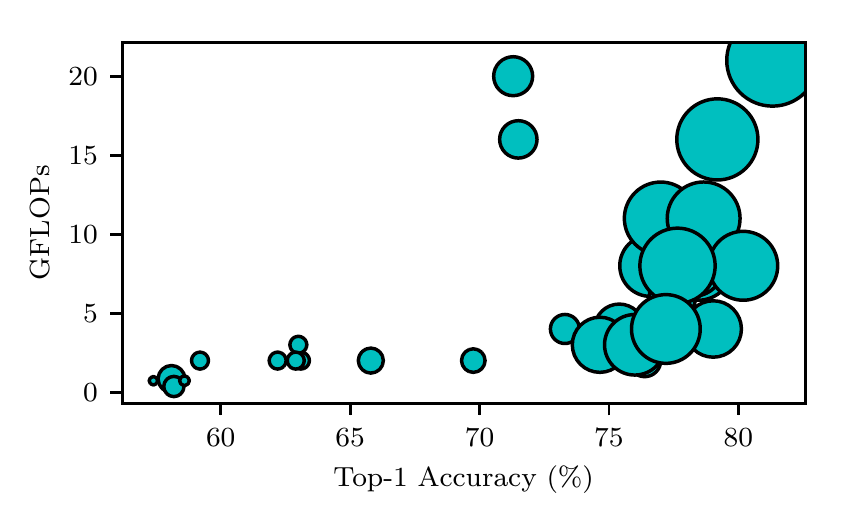}
\caption{Model accuracy and inference speed trade-offs of state-of-the-art CNNs (convolutional neural networks)~\cite{convnet_burden}. \textnormal{We approximate model inference speed with GFLOPs (giga floating point operations) and model size (circle size) assuming the same inference hardware. As shown, even for models with similar inference accuracy (around $78\%$), their inference time can be vastly different. \sysname leverages these models and makes accuracy and inference time trade-offs at runtime.}}
\label{fig:intro:cnnmodels} 
\end{figure}

To address the specific needs of mobile deep inference, we argue that systems need to take into consideration of mobile deep inference characteristics and automatically adapt inference execution to each inference request's accuracy and speed requirements. To combat the uncertainty in estimating both the inference network time and execution time in cloud-based inference, systems need to be able to  effectively explore and exploit  a set of models with minimal instrumentation overhead. 

Towards these two design goals, we propose \sysname, an algorithm for cloud-based inference that can adapt to network conditions by choosing the  deep learning model with the highest accuracy that will return results to the user within a request's SLA. \sysname is designed to optimize mobile inference accuracy based on the constraints of a given SLA, the time to transfer input data across the network, and the available models. 

More concretely, we enable a trade-off between inference speed and accuracy by, for each mobile inference request, choosing from a set of models that represent different execution time and accuracy tradeoffs, such as those shown in Figure~\ref{fig:intro:cnnmodels}. The key idea of \sysname is to match the time budget of an inference request to the deep learning model that is most likely to finish within the predefined SLA target and produce quality inference responses. 
We do this by first calculating an inference time budget for each request and then selecting a model from a subset of eligible models that are fast and reliable enough to meet this time budget.

Our hypothesis is that ModiPick's mobile-centric approach can significantly improve the accuracy of mobile inference requests while avoiding SLA violations. In evaluating our hypothesis, we make the following contributions.

\begin{itemize}
\item \textbf{Performance characterization of mobile deep inference.} We implement an image recognition Android application and a prototype serving system that manages a pool of convolutional neural network (CNN)  models. Our measurements show that sending inference requests to our cloud-based serving platform can be up to 3X slower in different mobile network conditions. In addition, we show that executing inference requests on the same model and cloud server combination incurs non-negligible time variation.  
\item \textbf{SLA-aware model selection policies.} \sysname uses a probabilistic-based approach when selecting the most suitable model for inference tasks based on user-defined SLA target. The key intuition behind this \emph{explore and exploit} algorithm is to account for the varying inference time budget caused by dynamic mobile environments and cloud inference time requirements. 
\item \textbf{Implementation and evaluation.} We implement \sysname as a python module that can be used by, and easily be integrated into existing deep learning serving systems~\cite{clipper,tensorflow_serving}. 
We conduct both microbenchmarks and end-to-end evaluations that demonstrate \sysname's ability to smoothly trade-off between inference accuracy and latency. Together with our extensive simulations that are seeded by empirical measurements, our results show that \sysname  achieves similar inference accuracy and increases SLA attainment by up to 88.5\% over greedy algorithms.
\end{itemize}

 \section{Motivation and Problem Statement}
\label{sec:bg}

Deep neural networks (DNNs) have become increasingly popular for embedding novel features into mobile applications. In particular, convolutional neural networks (CNNs) and recurrent neural networks (RNNs) have demonstrated high inference accuracy in handling image, video and audio data~\cite{alexnet, nasnet, 6638947}. State-of-the-art DNN models, with their accuracy-driven design, can contain millions of parameters and hundreds of layers, and therefore can be both computation and memory-intensive~\cite{resnet, nasnet, DBLP:journals/corr/SzegedyVISW15}. 

Mobile deep inference is defined as mobile applications use deep learning models to generate inference response for using in providing novel application features such as image recognition and speech recognition. Depending on where the deep learning model is executed, mobile deep inference can happen either directly \emph{on-device} or in \emph{cloud-based} servers~\cite{DBLP:journals/corr/Guo17a}. 
In this section, we first explain how on-device (\S\ref{sec:bg.on-device}) and cloud-based inference (\S\ref{sec:bg.in-cloud}) are used and discuss their pros and cons respectively. We then define the runtime model selection problem \sysname solves in \S\ref{sec:bg.problem-statement}.

\subsection{On-device inference and its limitations}
\label{sec:bg.on-device}

A number of deep learning frameworks, such as Caffe2~\cite{caffe2} and TensorFlow Lite~\cite{google_tensorflow:lite}, have started to support executing deep learning models directly on mobile devices. Specifically, these frameworks perform inference tasks using exported models that have trained on powerful servers. Even with optimizations in these software libraries, on-device inference can still be orders of magnitude slower than running inference on powerful servers (see Figure~\ref{bg:modelruntimes}). These large performance gaps are mainly due to constraints on mobile hardware, e.g., the lack of GPU or insufficient memory. The inference inefficiency is exacerbated when an application needs to load multiple models, such as chaining the execution of an OCR (optical character recognition) model and a text translation model~\cite{google_translate:ocr, google_translate:one_shot}, or needs higher accuracy models. To ensure mobile inference execution finishing in a reasonable time, mobile-specific models~\cite{DBLP:journals/corr/HowardZCKWWAA17, nasnet} often involves sacrificing inference accuracy and thus excludes complex application scenarios. 

\emph{Summary: Even though on-device inference is a plausible choice for simple tasks and newer mobile devices, it is less suitable for complex tasks or older mobile devices.}

\subsection{Cloud-based inference and its potential}
\label{sec:bg.in-cloud}

Instead of on-device execution, mobile applications can send their inference requests to cloud-hosted models. A number of model serving systems~\cite{clipper, tensorflow_serving, Gao:2018:LLR:3190508.3190541} have been proposed to manage different model versions that are used for executing inferences. These systems often focus on maximizing inference throughput by batching incoming requests, which may increase waiting time of some requests and consequently have negative impacts on end-to-end inference requests.   

To utilize such systems as model backends, mobile developers need to \emph{manually} specify the exact model to use through exposed API endpoints. This manual model selection fails to consider the impact of dynamic mobile network conditions, which can take up a significant portion of end-to-end inference time~\cite{MODI, Satyanarayanan:2009:CVC:1638591.1638731}. Such static development-time efforts can lead to mobile application developers either conservatively pick a faster model that delivers lower accuracy or risk SLA violations with a complex model that can take compound unexpectedly long network delay with a long inference time.

\emph{Summary: Cloud-based inference has the potential to support a plethora of application scenarios, simple or complex, and heterogeneous mobile devices, old or new. However, current mobile-agnostic serving platforms fall short in automatically adapting inference accuracy to varying time requirements of mobile inference requests.}

\subsection{Problem Statement}
\label{sec:bg.problem-statement}

\sysname seeks to execute mobile inference requests using the most suitable \emph{cloud-based} model to maximize inference accuracy within SLA target. 
 The key idea of \sysname is to mask the variability of dynamic mobile inference requirements by managing a pool of deep learning models that expose different accuracies and execution times. Specifically, these models (see Table~\ref{eval:model-summary} for an example distribution) provide \sysname the flexibility to adapt to the dynamic network and runtime environments experienced by each mobile inference request. 

In this work, we focus on answering the following two research questions: 
\textbf{(i)} how to handle dynamic mobile network by executing the inference task with the most suitable model that delivers optimized accuracy as SLA is allowed?
\textbf{(ii)} How to combat the unpredictability of cloud-based inference executions by judiciously exploiting potential model candidates?

\noindent{\bf System model.} We assume that mobile applications are packaged with an on-device model, but by default use the cloud-based inference APIs exposed by \sysname. For simplicity, we assume all inference requests are for image recognition tasks but  \sysname can be easily extended to other deep learning tasks by providing additional inference APIs.
\par
When sending an inference request, we assume that each mobile application include the request timestamp $T_{start}$ (all math symbols are defined in Table~\ref{tbl:symbols}), and the required SLA $T_{sla}$ in addition to the inference input data. We further assume that \sysname has access to a wide selection of models $\mathcal{M} = \{m_1, \cdots, m_k\}$ for the image recognition task and each model $m_i$ exposes a different level of accuracy $\bm{A}(m_i)$ and inference time $\bm{T}(m_i)$. Note, the models used are pretrained models that can be obtained from online model repositories~\cite{caffe:model_zoo, tensorflow_models} and derived using model optimization techniques~\cite{tensorRT, NetAdapt}. 

Here $\bm{A}(m_i)$ is calculated by dividing the correctly predicted inference requests by the total number of inference requests serviced by model $m_i$; and $\bm{T}(m_i)$ is defined as the time an inference server takes to generate an inference response. Note that $\bm{T}(m_i)$ includes both inference request wait time and execution time. In this paper, we assume that inference servers are well-provisioned and focus on making trade-offs between inference accuracy and execution time. That said, the problem of model servers provision is orthogonal to, but can be beneficial to our work.

 \section{\sysname Design} 
\label{sec:design}

This section describes how \sysname addresses mobile-specific key challenges using an adaptive and automatic model selection algorithm. 
We first motivate the need for maintaining a pool of cloud-based deep learning models in Section~\ref{subsec:needmodels}, followed by explaining how two baseline greedy algorithms fall short in Section~\ref{subsec:greedy}. 
We then describe \sysname's three-staged algorithm that automatically maximizes inference accuracy given variable inference time budgets and unpredictable cloud-based inference execution time in Section~\ref{subsec:modipickalgo}.

\subsection{The need of cloud-based multi-models hosting}
\label{subsec:needmodels}

In this paper, we consider mobile applications that provide novel features, which are supported by computational intensive deep learning models. These applications often have strict performance requirements and can only tolerate minimal accuracy degradation. 
As a result, despite recent device-specific model optimizations, hosting these models in the cloud is still a preferable approach for achieving scalability and consistentcy for heterogeneous mobile hardware~\cite{clipper, tensorflow_serving, velox}. 
In essence, the need for cloud-based models arise when these complex and accurate models still take very long time to execute on mobile devices (see Figure~\ref{bg:modelruntimes}). 
Moreover, having access to multiple cloud-based models can be very beneficial, as we explain below and demonstrate empirically in Section~\ref{sec:eval:benefits-of-multimodel-hosting}.

The key reasons that such deep learning powered mobile applications will benefit from having access to multiple model performance profiles at runtime lay at (1) the variable time budget for executing inference requests, (2) and the design goal of satisfying end-to-end response time SLA while optimizing inference accuracy.  Here, model performance profile refers to inference execution time when running a model on a particular cloud server. As shown in Figure~\ref{bg:modelruntimes}, the inference time of a specific model vary significantly depends on cloud server types. In essence, we can dynamically select which model and sever combination to use given the inference time budget. To cater to applications with different accuracy requirements, maintaining a pool of deep learning models that expose different accuracy and computation complexity can provide greater flexibility. 

When a mobile application sends an inference request to a cloud-based model, the end-to-end inference time is impacted by input transfer time, input preprocessing time, and inference execution time.
For image recognition tasks, input preprocessing time is often negligible even between recent mobile hardware and much older devices~\cite{DBLP:journals/corr/Guo17a}.
However, as explained in Section~\ref{sec:intro}, the time to transfer input can vary significantly depending on the input data size and network connection.
As a result, cloud-based model servers have to adhere to a variable time budget in order to avoid SLA violations. 

However, current serving platforms often statically choose a model with an acceptable accuracy that can finishes its inference execution within an upper bound.
This bound can be calculated by taking into account the worst-case scenarios of input transfer and preprocessing time. Consequently, such single model hosting is too conservative and ignores opportunities to achieve higher inference accuracy for scenarios where a fast network connection allows for a higher execution time budget. In summary, relying on a single model to handle mobile deep inference is not optimal as there are no ``one-size-fits-all'' models.  

Instead, we argue the need for maintaining a pool of deep learning models that expose different accuracy and speed tradeoffs, as shown in Figure~\ref{fig:intro:cnnmodels}. Our proposed cloud-based multi-model inference service has two major advantages. \emph{First}, it provides the flexibility for \sysname to automatically adapt its model selection to varying inference execution budget, for each incoming inference request. Therefore, \sysname is able to strive for higher inference accuracy within predefined SLA target. \emph{Second}, models with increasing complexities can satisfy a given execution budget by running on more powerful servers. Therefore, \sysname can deliver the same inference response with reduced cloud cost. In this paper, we focus on the SLA-aware accuracy-optimization that are enabled by multi-model hosting and leave the cost-optimization as future work.

\begin{table}[t]
\centering
\resizebox{\columnwidth}{!}{
\begin{tabular}{l|l}
\textbf{Symbol}          & \textbf{Meaning}                                            \\ \hline
\rowcolor{Gray}
$T_{sla}$       & Response time SLA.                                 \\
\rowcolor{Gray}
$T_{start}$     & Start time of mobile inference requests.           \\
\rowcolor{Gray}
$T_{threshold}$ & Confidence threshold of inference performance.     \\
\rowcolor{Gray}
$T_{budget}$    & The remaining time for inference execution.        \\
\rowcolor{Gray}
$T_{nw}$        & Network time of inference request and response.    \\
\rowcolor{Gray}
$T_{input}$     & Time to send inference request.                    \\
\rowcolor{Gray}
$T_{output}$    & Time to send inference response.                   \\
\rowcolor{Gray}
$T_{D}$         & Expected on-device inference time.                 \\
\rowcolor{Gray}
$T_{U}$         & The hard time limit for model exploration.         \\
\rowcolor{Gray}
$T_{L}$         & The soft time limit for model exploration.         \\
\rowcolor{Gray}
$T_{E}$ & The exploration range. \\
\rowcolor{Gray}
$T_{R}$         & The range for model exploration.                   \\
$K$             & The total number of models.                        \\
$M_E$           & The exploration set of models.                     \\
$\mu(m)$        & Average inference time of model $m$.               \\
$\sigma(m)$     & Standard deviation of inference time of model $m$. \\
$\bm{Pr}(m)$    & Probability of model $m$ for performing inference. \\
$\bm{A}(m)$     & Accuracy of model $m$.                             \\
$\bm{T}(m)$     & Inference time of model $m$.                       \\
$\bm{U}(m)$     & Utility of model $m$.                             
\end{tabular}
}
\caption{Summary of symbols and their meanings. \textnormal{Symbols are categorized as inference-related (shaded) and model-related ones. For inference-related symbols, all but $T_{sla}$ symbols are in regarding to a particular inference request $R_j$ from a specific mobile device $D_i$. But for simplicity, we omit the representation of inference requests and devices throughout the paper.}}
\label{tbl:symbols}
\end{table}

\subsection{Baseline greedy approaches} 
\label{subsec:greedy}

Provided the feasibility and benefits of hosting multiple deep learning models for a given type of inference tasks, the next step is to determine which model to use for any given inference request. \sysname's goal is to select the most appropriate model by considering both dynamic inference budget and the cloud-based inference time variation. 
In this section, we first present two greedy approaches (used as baselines in Section~\ref{sec:eval}) and identify their respective limitations. In this section~\ref{subsec:modipickalgo}, we introduce our probability-based model selection that aims to achieve high inference accuracy without violating SLA. 

\subsubsection{Static greedy model selection}

One approach to selecting a model is to study the accuracy and latency trade-offs of all models at \emph{development time}.
These models can then be sorted in descending order of accuracy, and the model with the highest accuracy that has a response time less than the predefined SLA $T_{sla}$ will be selected. 
That is, we pick the first model $M_k$ with its average inference execution $\mu({M_k}) \le T_{sla}$.
While this approach is straightforward, it requires mobile application developers to specify the selected model during development time and is therefore inherently unable to cope with the dynamic environment experienced by mobile devices.
Moreover, ignoring the variability of mobile network connections can potentially lead to a large number of SLA violations that are unacceptable in user-facing mobile applications.

\subsubsection{Dynamic greedy model selection}
An alternative, but also na\"ive, approach is to decide which model to use for executing inference at runtime. 
In order to make the decision without violating application-specified SLA $T_{sla}$, this approach estimates the remaining time $T_{budget}$ that an inference request has to finish execution for each request. This time budget is calculated by taking the difference between $T_{sla}$ and the network transfer time $T_{nw}$. $T_{nw}$ can be estimated conservatively with $2*T_{input}$ where $T_{input}$ denotes the time taken to send input data from the mobile device to the inference server. In most cases, we could expect $T_{input} \geq T_{output}$ given inference requests, e.g., images, are often larger than inference responses. To summarize, $T_{budget}$ can be calculated as: 

\begin{equation}
T_{budget}=T_{sla} - 2*T_{input} \label{eq:budget}
\end{equation}

Given the time budget, this approach sorts the models in descending order of prediction accuracy and picks the first model $M_k$ with its average inference execution $\mu({M_k}) \le T_{budget}$. Note, the key differences between this approach and the previously described \emph{static greedy} is \emph{when} the model selection is made and \emph{how} the model is selected.  

Although this satisfies our goal of optimizing inference accuracy, this dynamic approach is not resilient to scenarios when we do not have accurate performance estimates for hosted models.
Such situations can arise when a model is first deployed to the inference server or when inference execution time vary vastly due to overloaded servers~\cite{8298516} or performance interference of co-located tenants, as demonstrated in Figure~\ref{bg:modelruntimes}. 

Given the uncertainty of mobile network and cloud performance, we aim to maximize the prediction accuracy by balancing the opportunities to exploiting our current knowledges and exploring other potentially good candidate models~\cite{Yogeswaran2012}. Without exploration, high-accuracy models that happen to incur abnormally high inference time can be omitted from any future selections. We next depict our accuracy-driven probabilistic model selection that is centered around this key insight.

\subsection{Accuracy-driven dynamic model selection} 
\label{subsec:modipickalgo}

Selecting the best deep learning model to execute mobile inference requests can be challenging given the uncertainty of inference time budgets and unpredictable cloud-based inference performance. The key insight of our accuracy-driven model selection algorithm is that by \emph{explicitly} considering such uncertainty, mobile inferences on average increase inference speed and accuracy, without incurring significant performance profiling overhead. 

\begin{figure}[t]
\begin{center}
\includegraphics[width=0.45\textwidth]{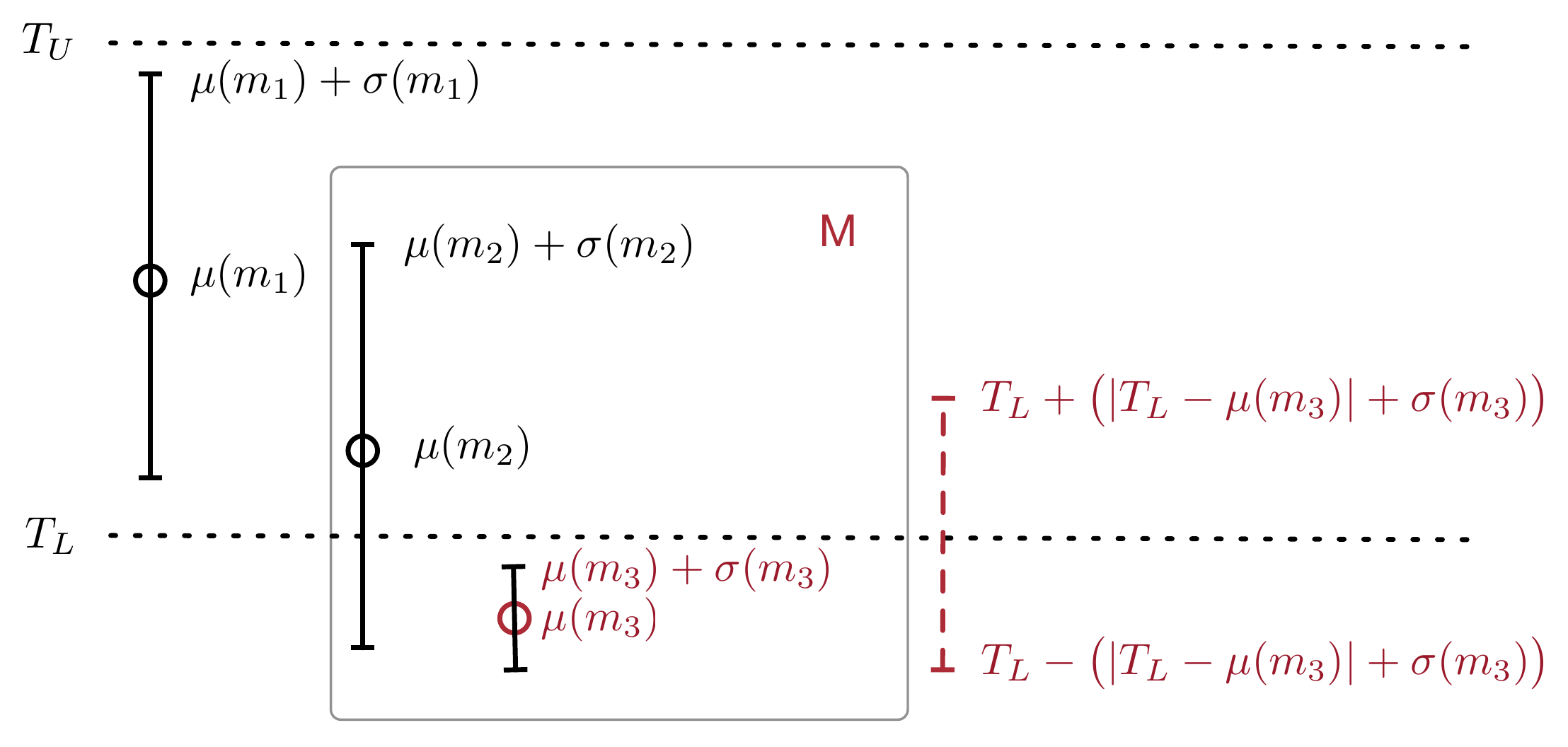}
\caption{Walkthrough of our accuracy-driven model selection. \textnormal{\sysname builds performance profile for each model by tracking average inference time $\mu(m)$ and standard deviation $\sigma(m)$. Assuming $\bm{A}(m_3) > \bm{A}(m_1) > \bm{A}(m_2)$. 
To select the model for current inference request, \sysname will first selects $m_3$ as a base model, and then calculate the exploration range $T_E$, illustrated as the dashed red line. \sysname then constructs the exploration set $M_E$ that contains $m_2$ and $m_3$. Last ,\sysname chooses between $m_2$ and $m_3$ with a probability that takes into account of both model accuracy and inference time variability.}}
\label{fig:design.model.selection}
\end{center}
\end{figure}

In order to effectively explore all potentially high-accuracy models without violating SLA, we define a threshold $T_{threshold}$ which indicates how uncertain we are about the model performance profiles. The larger the value of $T_{threshold}$, the more outdated the inference performance profiles. We then expand the notation of time budget $T_{budget}$ to a range $T_R = [T_U, T_L]$, where  $T_U=T_{budget}$ and $T_L  = T_U - T_{threshold}$ (see Figure~\ref{fig:design.model.selection}).  Intuitively, $T_U$ represents the maximum amount of time that \sysname can use for generating an inference response without risking SLA violations. We refer to $T_U$ as the \emph{hard time limit}. On the other hand, $T_L$ is referred to as the \emph{soft time limit} and provides \sysname the flexibility to explore a subset of high-accuracy models $M_{E}$ that exhibit different execution time $\{\bm{T}(m) | \forall m \in M_{E}\}$.

Currently, $T_{threshold}$ is configured by \sysname's users, e.g., mobile application developer, as any values in the range of $[0, T_{D}]$, where $T_{D}$ represents the expected on-device inference time for a mobile application. However, \sysname could also dynamically adjust $T_{threshold}$ based on its confidences of model performance profiles and will be explored as part of future work. We choose to bound $T_{threshold}$ this way because (1) we want to restrict the set of the candidate models $M_{E}$ for exploring, (2) and it can mitigate the undesirable behavior of starting on-device inference prematurely when cloud-based inference can finish without violating the SLA.

Next, we describe in detail how \sysname utilizes both the model performance profiles and the time budget range $T_R$ to first pick a base model, then construct a set of eligible models $M_E$ that is worth exploring, and last probabilistically select the model for executing the inference request. Our three-staged algorithm is designed to gradually improve our estimation of model performance profiles without incurring additional profiling overhead. In addition, if we are under time pressure to select models, our algorithm could be stopped any time after the first stage and will still select a quality model for performing the inference. 
In Figure~\ref{fig:design.model.selection} we provide an example walkthrough of how \sysname uses our accuracy-driven model selection algorithm to select the best model probabilistically in order to combat both the uncertainty of mobile network and the inaccuracy of model performance in the cloud environment.  

\subsubsection{\textbf{Stage one} greedily picking the baseline model} 
In this stage, \sysname takes all the existing models and selects a base model $m_j$ based on the optimization formulation in Equation~\ref{eq:stage1}.

\begin{equation}
\begin{aligned}
& \underset{j}{\text{maximize}}
& & \bm{A} (m_j)  \\
& \text{subject to} && \mu(m_j) + \sigma(m_j) < T_U , \; j = 1\ldots K. \\
&&& \mu(m_j) - \sigma(m_j) < T_L , \; j = 1 \ldots K. \\
\end{aligned}
\label{eq:stage1} 
\end{equation}

The high level idea is to select the most accurate model (objective function) that are likely to finish execution within specified SLA target (first constraint) without triggering on-device inference (second constraint).  
Given that cloud-based inference execution might experience performance fluctuations that lead to a wider inference execution distribution~\cite{Sharma:2016:CVM:2988336.2988337,8298516}, we take into account of the standard deviation of model inference time and only select models that satisfy both the soft time limit $T_L$ and the hard time limit $T_U$. By doing so, the selected models are of high accuracy and are very likely to finish execution within specified SLA. Note that when $T_{threshold}=0$ and all models have very tight inference distribution, i.e., $\sigma(m_j) < \epsilon, \forall j$ where $\epsilon$ is small, the base model selected by Equation~(\ref{eq:stage1}) will be the same as the one by Equation~(\ref{eq:budget}). In the example walkthrough in Figure~\ref{fig:design.model.selection}, \sysname will select model $m_3$ as the base model.

Due to the variability of the network, there will occasionally be cases in which there are no available base models that satisfy Equation~(\ref{eq:stage1}).
In these situations, \sysname will choose the model with the lowest average execution time $\mu(m_j)$ in order to provide a best-effort at SLA attainment.

\subsubsection{\textbf{Stage two} optimistically constructing the eligible model set}

While Figure~\ref{bg:modelruntimes} shows that cloud-based inference has largely stable execution times, there are cases in which these execution times are orders of magnitude worse or may be unknown.
For instance, if a model happens to incur a very long inference time due to sudden workload spikes on the co-located cloud tenants and becomes 
ineligible based on Equation~(\ref{eq:stage1}), a non-probabilistic system would never pick this model for execution again even though it might provide competitive accuracy. 
Another example is when a new model is submitted to \sysname and there is not sufficient performance history related to it in order to make a determination regarding expected inference time. In this case the model needs to be run until an accuracy and performance profile can be generated.

To take the above described scenarios into consideration, we leverage the basic idea of exploiting and exploration~\cite{Yogeswaran2012} and 
use the base model as the anchor to construct a subset of eligible models $M_E$. This subset of models $M_E$ satisfy our performance goal regarding accuracy and inference time,
while providing the opportunity for \sysname to improve upon their performance profiles.

The high level idea is to first identify a reasonable range $T_E$ for exploration and then only select models that satisfy $\mu(m) \in T_E$. 
Intuitively, the wider the range, the more models are likely to fall within the range. It is also obvious that we don't want $T_U$ to fall in such ranges 
because the increased chances of SLA violations. Therefore, we construct the range $T_E$ by centering around the soft time limit $T_L$ and expand 
the acceptable inference time values in both directions with the distance between the base model and $T_L$, augmented with the standard deviation of the base model inference time. 
In addition, in order to minimize SLA violations, a model $m$ is only chosen to put in the eligible model set $M_E$ when $\mu(m) + \sigma(m) < T_U$. 
In Figure~\ref{fig:design.model.selection}, only model $m_2$ and $m_3$ are marked as the members of $M_E$.

\subsubsection{\textbf{Stage three} opportunistically selecting the inference model} 
Once \sysname constructs the exploration set of models $M_E$, its ultimate goal is to select an inference model $m_j$ that balances the exploration reward and the risk of SLA violations. 
To do so, we resort to a probabilistic approach that assigns each model a probability based on its utility and selects a model based on its corresponding probability. For each model $m$, its probability $\bm{Pr}(m)$ represents \sysname's understanding of model accuracy $\bm{A}(m)$ and the gaps of violating either hard or soft time limit.  To derive such probability values, we first denote each model's utility using $\bm{U}(m)$. 

\begin{equation}
\bm{U}(m) = \bm{A}(m) \frac{T_U - \big(\mu(m) + \sigma(m)\big)}{| T_L - \mu(m) |}\label{eq:utility}
\end{equation}

In Equation~\ref{eq:utility}, the numerator $T_U - \big(\mu(m) + \sigma(m)\big)$ is always positive based on our algorithm stage two that selects the exploration set of models $M_E$. If a model $m$ has a wide range of inference time with larger standard deviation, we try to avoid assigning higher probability for such models. Similarly, if a model $m$ on average has a larger absolute difference to the soft time limit $T_L$, indicating a lower confidence of its performance profile, 
we want to avoid such models as well. Overall, Equation~\ref{eq:utility} assigns a model $m$ with a higher probability if the model has a higher accuracy, less likely to violate the SLA  and has an up-to-date and accurate inference execution time.
When picking the model for inference, \sysname chooses the model from the exploration set  $M_E$ based on its normalized probability $\bm{Pr}(m)$ in Equation~\ref{eq:prob}. 

\begin{equation}
\bm{Pr}(m) = \frac{1}{\sum\limits_{n \in \bm{M_E}} \bm{U}(n)} \bm{U}(m)\label{eq:prob}
\end{equation}

\noindent \textbf{Practical considerations: }\emph{maintaining cloud-based model performance profiles.} In order to maintain an accurate and update-to-date model inference execution performance when using cloud servers, we periodically re-evaluate the inference latency for each model, especially for models that have not been selected recently. Unlike popular models, these ``cold" models could not leverage our runtime measurement to effectively update the performance profiles and might be completely ruled out due to past poor performance. In addition to obtaining performance profiles , we also use an exponential weighted moving average to calculate $\mu(m)$ and $\sigma(m)$ for each model $m$ to combat the performance fluctuations over time.

 \section{Experimental Evaluation}
\label{sec:eval}

Our evaluation goal is to quantify the effectiveness of \sysname, in dynamically selecting the most appropriate deep learning model to optimize inference accuracy while avoiding SLA violations. 
We first present an end-to-end experiment that demonstrates \sysname's ability to improve inference accuracy by adapting its model selection decisions as SLA target increases, in Section~\ref{sec:eval:prototype}. 
We then compare the performance of \sysname and baseline greedy algorithms (refer to Section~\ref{subsec:greedy}) in terms of inference time and accuracy, under different SLA targets and mobile network conditions, in Section~\ref{sec:eval:benefits-of-multimodel-hosting} and Section~\ref{eval.subsection.dynamic_network_conditions}. We conduct a comprehensive analysis that attributes the performance gain to the ability to utilizing a diverse set of models. Lastly, we investigate the benefits brought by \sysname's accuracy-driven dynamic model selection algorithm in Section~\ref{subsec:decomposing_benefits}. Our experiments are performed using our prototype systems while our simulations allow us to study the benefits in a scalable way, e.g., evaluating through a large number of mobile inference requests with different SLA and network condition combination, as well as using a large number of deep learning models.

\noindent \textbf{Prototype setup.} Our prototype serving system runs on a well-provisioned Amazon EC2 \texttt{p2.xlarge} server in the Virginia data center and manages two deep learning models, \texttt{MobileNetV1 0.25} and \texttt{InceptionV3}, through our \sysname algorithm.
These two models are retrained on a smaller dataset and can deliver inference accuracy of 88.9\% and 94.3\% respectively.
We choose these two models to better demonstrate the trade-offs that occur when using two models.The serving system is first warmed up by executing both models using 1000 sample inference requests to allow \sysname to establish both models' performance profiles (see Figure~\ref{bg:modelruntimes}).
To send mobile inference requests, we use our image recognition Android application running on a reasonably powerful mobile device (MotoX) via a campus WiFi. For each SLA target, our mobile application sends 1000 inference requests to the serving system and measure both the accuracy and the inference time.  

\noindent \textbf{Simulation setup.} In our simulations, we leverage a range of models, summarized in Table~\ref{eval:model-summary}\cite{tensorflow_models, DBLP:journals/corr/HowardZCKWWAA17, DBLP:journals/corr/SzegedyVISW15, NIPS2012_4824, DBLP:journals/corr/IandolaMAHDK16, DBLP:journals/corr/SimonyanZ14a, resnet}, that expose different accuracy and inference time trade-offs. We empirically measured the inference time distributions of models using an EC2 \texttt{p2.xlarge} GPU-accelerated server over 1,000 inference executions. Model accuracies are obtained from the original publications unless otherwise noted. 
We simulate the mobile network profiles based on empirical measurements of network time (average = 57.87ms, std = 30.78ms) to send an inference image (330KB) from mobile app to a Virginia-based EC2 server via our campus WiFi. For each simulation, we generate 10,000 inference requests with a predefined SLA target and record the model selected by \sysname (and baseline algorithms) and relevant performance metrics. We repeat each simulation for different SLA target and network profiles combination.

\begin{table}[t]
\resizebox{\columnwidth}{!}{
\begin{tabular}{|l|l|l|l|}
\hline
\textbf{Model Name}     & \textbf{Top-1 Accuracy (\%)} & \textbf{Inference  Avg. (ms)} & \textbf{Inference Std. (ms)} \\ \hline
SqueezeNet              & 49.0                    & 4.91 &  0.06        \\ \hline
MobileNetV1 0.25        & 49.7                    & 3.21 &  0.08      \\ \hline
MobileNetV1 0.5         & 63.2                    & 4.21 &  0.06         \\ \hline
DenseNet                & 64.2                    & 25.49 &  0.14        \\ \hline
MobileNetV1 0.75        & 68.3                    & 4.67 &  0.07         \\ \hline
MobileNetV1 1.0         & 71.0                    & 5.43 &  0.11         \\ \hline
NasNet Mobile           & 73.9                    & 21.18 &  0.17        \\ \hline
InceptionResNetV2       & 77.5                    & 50.85 &  0.33        \\ \hline
InceptionV3             & 77.9                    & 31.11 &  0.19       \\ \hline
InceptionV4             & 80.1                    & 59.21 &  0.22        \\ \hline
NasNet Large            & 82.6                    & 112.61 & 0.36       \\ \hline
NasNet Fictional* 	& 50 & 112.61 & 0.36       \\ \hline
\end{tabular}
}
\caption{Summaries of model statistics through empirical measurement. \textnormal{Models are sorted based on their top-1 accuracy which is defined as the percentage of correctly labeled test images using only the most probable label. We measure the average inference time (third column) and standard deviation (last column) for each model running on an EC2 \texttt{p2.xlarge} GPU server. We use these state-of-the-art models in simulations to study \sysname's effectiveness in trading-off inference accuracy and time. Note \texttt{NasNet Fictional} is a made-up model based on \texttt{NasNet Large} and is only used in Section~\ref{subsec:decomposing_benefits}.}}
\label{eval:model-summary}
\end{table}

\begin{figure}[t]
\centering
\includegraphics[width=\columnwidth]{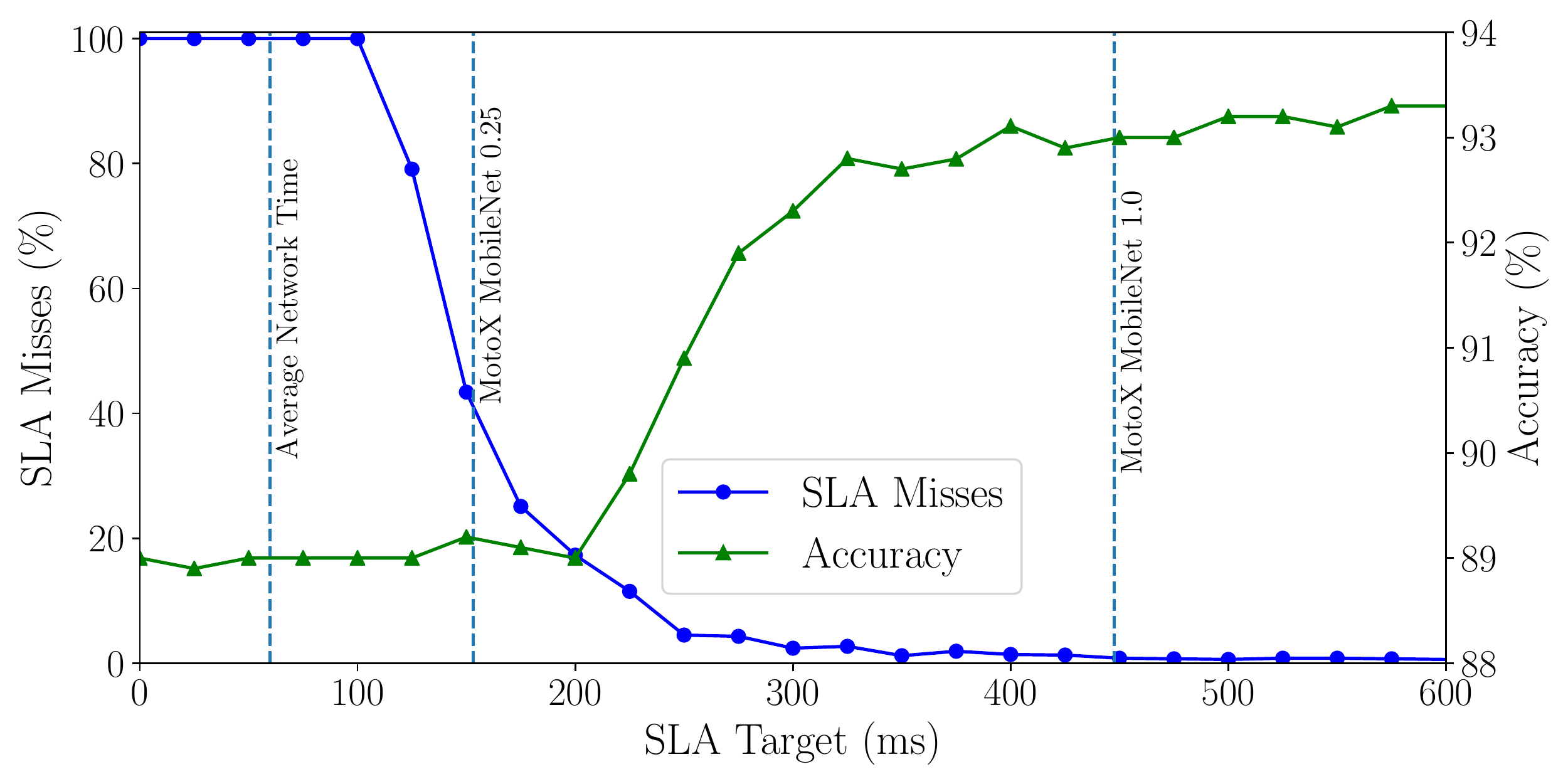}
\caption{End-to-end performance of \sysname with prototype systems. \textnormal{\sysname is able to automatically transition between models with different accuracy and inference runtime.  As the SLA target increases beyond the network latency \sysname can begin returning results using a low-latency model.  As the SLA increases further it can begin using a more accurate model for inference, increasing the accuracy while continuing to decrease SLA violations.} }
\label{eval:subfig:prototype:trade-off}
\end{figure}

\begin{figure}[t]

    \begin{subfigure}{0.45\textwidth}
    \centering
    \includegraphics[height=0.5\columnwidth]{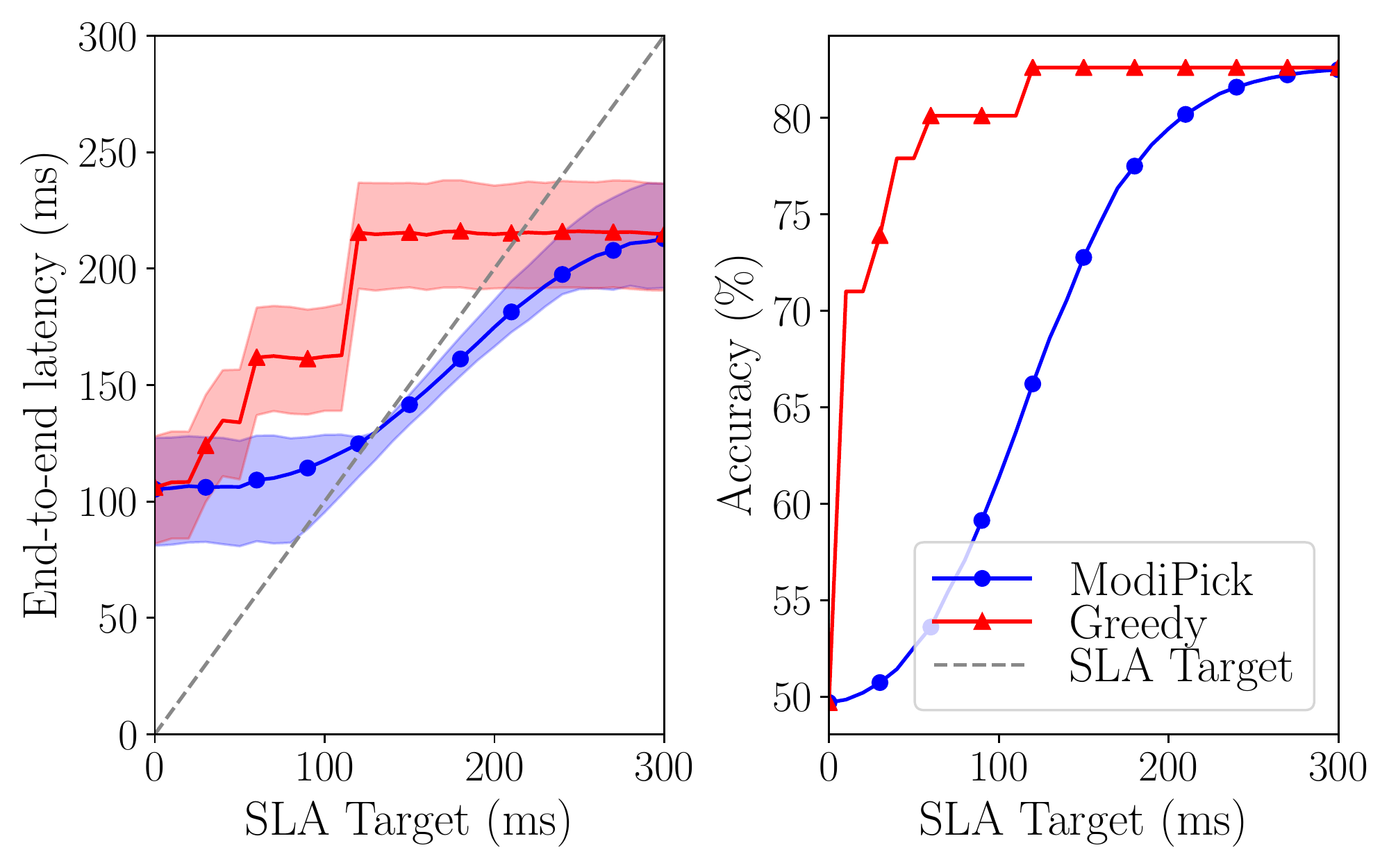}
        \caption{\textnormal{Comparison of average end-to-end latency (shaded with one standard deviation) and accuracy. \sysname is able to keep track of the SLA target when SLA $\geq$100ms while the greedy approach fails to do so. \sysname is able to improve achieved model accuracy \emph{safely} as the SLA target increases. Note that the static greed model experiences end-to-end latency variation due to the network latency since it cannot correct for it at runtime. }}
    \label{eval:subfig:two-models:compare} 
    \end{subfigure}
        
    \hfill
    
    \begin{subfigure}{0.45\textwidth}
    \centering
    \includegraphics[height=0.5\columnwidth]{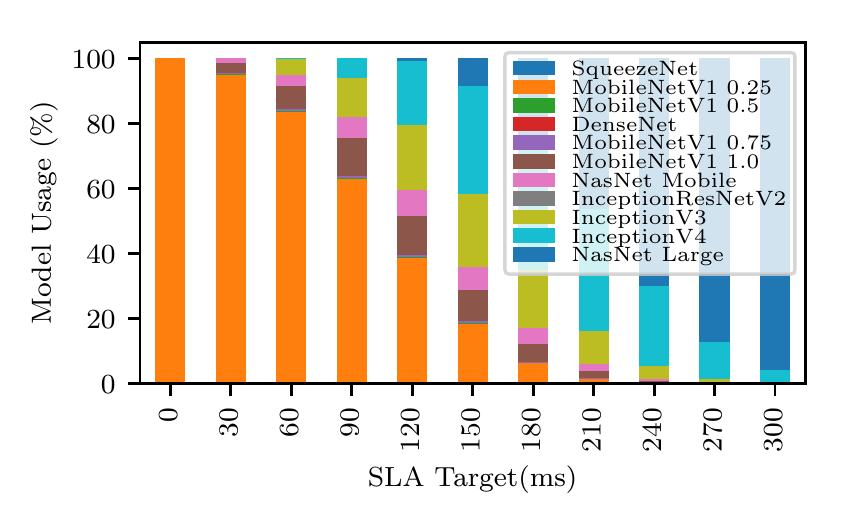}
        \caption{\textnormal{Illustration of the models used by \sysname to enable smooth trade-off between latency and accuracy. The use of a diverse set of models allows \sysname to minimize SLA violations while providing highly accurate inference results.}}
    \label{eval:subfig:two-models:model-usage} 
    \end{subfigure}
    
    \caption{Comparison of \sysname to the \emph{static greedy} algorithm described in Section~\ref{subsec:greedy}. \textnormal{For each SLA target, we simulated 10,000 inference requests and recorded the inference time incurred by both the greedy and \sysname.}} 
    \label{eval:fig:two-models}
\end{figure}

\begin{figure}[t]
    \centering
    \includegraphics[width=0.5\textwidth]{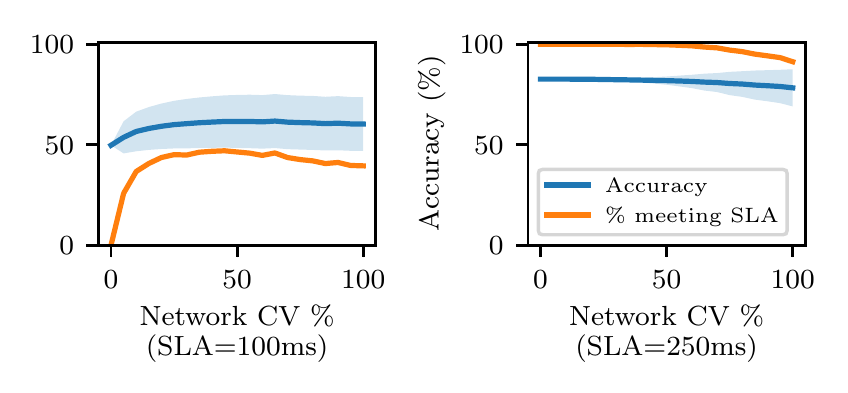}
        \caption{Effective accuracy of \sysname at different levels of CV with standard deviation shown.  
                \textnormal{The initial low level of SLA attainment is due to the fact that the network time is initially 100ms, with a standard deviation of 0ms.  As the variability of the network increases \sysname can take advantage of the range of models available to it to quickly improve accuracy and SLA attainment.  Similarly, at a higher SLA, \sysname can achieve high accuracy until the network variability becomes too much at which point it begins having to use lower latency models.}}
    \label{eval:fig:CV:accuracy}   
\end{figure}

 \begin{figure*}[t]
    \centering
    \includegraphics[width=\textwidth]{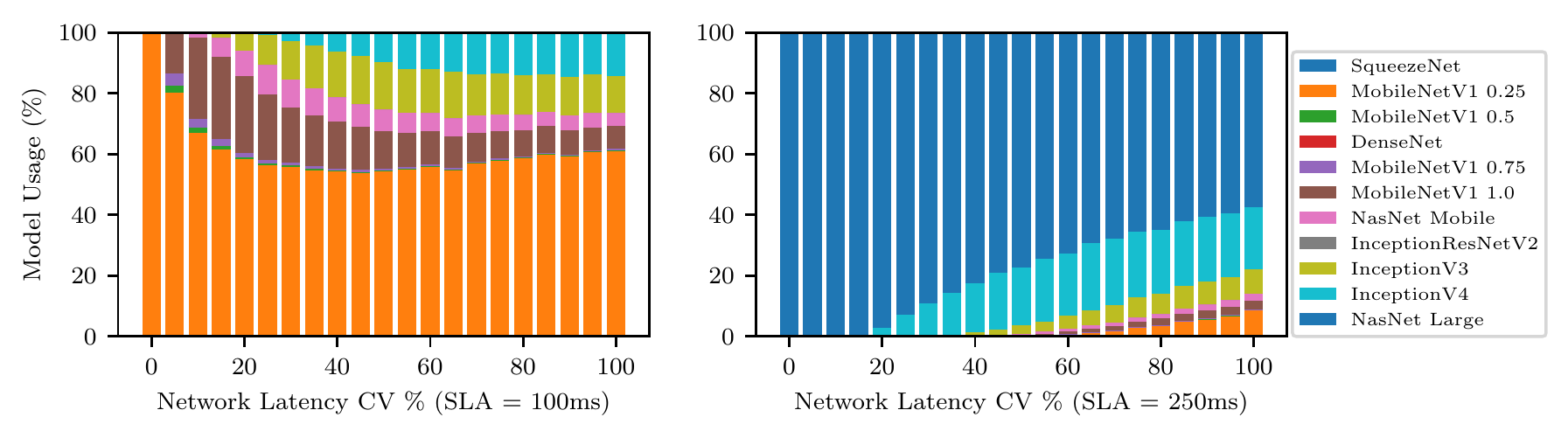}
    \caption{ Model usage vs. network latency (CV) shown at two different SLA points.  \textnormal{When there is a reliable network (i.e. low CV) single models dominate as all requests have the same inference time budget.  As the network becomes increasingly volatile a wider range of models, including both higher and lower latency models, is used to meet the SLA.} }
    \label{eval:fig:CV:model-usage}
\end{figure*}

\subsection{Prototype evaluation}
\label{sec:eval:prototype} 

We first demonstrate \sysname's overall effectiveness with an end-to-end experiment using our prototype serving system and an Android application running on MotoX.
The mobile device is connected to our campus WiFi which has an average network time of $63$ms over the course of the test.
For each mobile inference request, the image recognition mobile app will first preprocess the image (to 330KB) and then send the image together with other relevant meta data, e.g., SLA target and request timestamps, to our Virginia-based serving system. 

In Figure~\ref{eval:subfig:prototype:trade-off}, we plot the percentage of inference requests that violate SLA (left y-axis) and the percentage of inference requests that are correctly classified (right y-axis) for different SLA targets. We further annotate the figure with three important timelines: average network time, on-device inference time with \texttt{MobileNetV1 0.25} and on-device inference time with \texttt{MobileNetV1 1.0} (from left to right) to better illustrate \sysname performance. 

As we can see, \sysname is able to gradually reduce the percentage of SLA misses as the SLA target increases. In particular, we start to observe reduction in the number of SLA violations and improved inference accuracy once the SLA target is larger than $115$ ms. This is due to \sysname recognizing that the time budget is extremely small and beginning to choose a small model, \texttt{MobileNetV1 0.25}, that can quickly return inference results. As the SLA increases further, the overall inference accuracy begins to improve but still exhibits some variation. The improved accuracy is due to \sysname identifying the increased time budget and beginning to use the more accurate \texttt{InceptionV3} model while the continuing variation in accuracy is due to \sysname accounts for network variability and occasionally chooses \texttt{MobileNetV1 0.25}. 

\emph{Summary: \sysname is able to adapt its model selection with the goal to minimize SLA violations while improving inference accuracy, even when SLA target is set to be as low as executing a mobile-optimized model on-device.}

\subsection{Benefits over static greedy model selection}
\label{sec:eval:benefits-of-multimodel-hosting} 

To examine \sysname's ability to handle the trade-offs between inference latency and accuracy, we run a simulation with first eleven deep learning models  outlined in Table~\ref{eval:model-summary} (i.e. all models except \texttt{NasNet Fictional}).
We simulate mobile network based upon our empirically measured campus WiFi network (see Figure~\ref{fig:background:in-cloud}).
Note that this network has the lowest network latency and standard deviation among our measured network.
We further examine the efficiency of \sysname under variable network conditions in Section~\ref{eval.subsection.dynamic_network_conditions}. 
We compare \sysname to the \emph{static greedy} algorithm (see Section~\ref{subsec:greedy}) that always chooses the most accurate model for a given SLA.  

In Figure~\ref{eval:subfig:two-models:compare}, we plot the average end-to-end inference time (left) and inference accuracy achieved by these two algorithms.
This figure shows that \sysname consistently achieves up to 42\% lower inference latency, compared to \emph{static greedy}. Moreover, \sysname can operate under a much more stringent SLA target ($\mathtt{\sim}115ms$) while \emph{static greedy} continues to incur SLA violations until SLA target is more than 200ms. The key reason is because \sysname is able to effectively trade-off accuracy and inference time by choosing from a diverse set of models (see Figure~\ref{eval:subfig:two-models:model-usage}). Consequently, \sysname has an accuracy of 68\% (on par to using MobileNetV1 0.75 which can take 2.9x more time running on mobile devices) under low SLA target ($\mathtt{\sim}115ms$), but is able to match accuracy achieved by \emph{static greedy} when SLA target is higher. Note that even though \emph{static greedy} achieves up to 12\% higher accuracy, it does so by sacrificing inference latency. 

In Figure~\ref{eval:subfig:two-models:model-usage}, we further analyze \sysname performance by looking at its model usage patterns under different SLA targets. At very low SLA target (<30ms), \sysname aggressively chooses the fastest model \texttt{MobileNetV1 0.25} since none of the managed models satisfy Equation~\ref{eq:stage1}. As the SLA target increases, \sysname explores more accurate but slower models than \texttt{MobileNetV1 0.25}. There are two key observations: (1) \sysname is effective in picking the more appropriate model to increase accuracy while staying safely within SLA target. For example, \texttt{InceptionResNetV2} is never selected by \sysname because better alternatives \texttt{InceptionV3} for lower SLA target and \texttt{InceptionV4} for higher SLA target exist. (2) \sysname faithfully explores eligible models and is able to ``converge" to the most accurate model when SLA target is sufficiently large.

\emph{Summary: \sysname outperforms \emph{static greedy} with up to 43\% end-to-end latency reduction, while is able to keep up with accuracy with SLA budget is larger than $250$ms. The key reason is because \sysname is able to adapt its model selection by considering both the SLA target and network transfer time, while \emph{static greedy} naively selects the most accurate model.}

\subsection{Adaptiveness to dynamic mobile network conditions}
\label{eval.subsection.dynamic_network_conditions}

One of the key reasons that \sysname can meet any given SLA targets while optimizing accuracy whenever possible is its ability to adapt to variations in network transfer time.
To closely examine how \sysname copes with these variations we simulate network profiles with increasing variability. Specifically, we fix the average network latency to be $50$ms, and vary CV (Coefficient of Variation) from 0\% to 100\%. Here CV is defined as the ratio between standard deviation and average, and a CV of 0\% indicates a very stable network condition while a CV of 100\% means the network latency distribution is dispersed with its standard deviation equals to its average. As a point of reference, our measured Campus WiFi network has a CV of 74\%. 

In Figure~\ref{eval:fig:CV:accuracy}, we plot the average inference accuracy and SLA attainment (percentage of requests that meet the SLA target) achieved by \sysname. For low SLA target (100ms), when the network is relatively stable, \sysname has a SLA attainment of less than 50\%. As the network condition becomes more variable, \sysname is able to increase the inference accuracy up to 10\% gradually while maintaining the SLA attainment. However, it is important to note that \sysname performs as expected by choosing the fastest possible model, i.e., \texttt{MobileNetV1 0.25}. We note that, to satisfy such stringent SLA targets with high network latency, alternative approaches such as provisioning inference servers near network edge or executing latency-optimized \texttt{MobileNetV1 0.25} on-device are generally preferable. On the contrary, when given a reasonable SLA target, e.g., slightly more than average network latency plus the time to execute the most accurate model \texttt{NasNet Large}, \sysname is resistant to network variation well with high SLA attainment    and maintains an accuracy around 80\% (slightly less than \texttt{NasNet Large}). 

Figure~\ref{eval:fig:CV:model-usage} demonstrates the percentage of selected models with increasing network variability for both a very slow SLA target (100ms) and a reasonable SLA target (250ms). There are there important trends to observe: (1) as the network becomes more variable (larger value of CV), \sysname  matches the network variability with a subset of faster models. This is because as \sysname becomes less certain about the inference request time budgets, \sysname starts to explore for models with higher inference accuracy. (2) The probability of exploring different eligible models is proportional to the SLA target and network variability. Faster models, such as those in the MobileNetV1 family, are used as a basis for low SLA target while the most accurate model, i.e., NasNet Large, is used for a reasonable SLA target. (2) For different SLA targets, \sysname chooses to explore different subset of models, e.g., for low SLA target, \texttt{MobileNetV1 family} and \texttt{NasNet Mobile} and \texttt{InceptionV3}, for a reasonable SLA target, \texttt{NasNet Large}, \texttt{InceptionV4}, \texttt{InceptionV3}, and \texttt{NasNet Mobile}.

Figure~\ref{eval:fig:CV:model-usage} shows the models chosen when varying CV of network time for SLAs of 100ms and 250ms.
The SLA of 100ms is the RTT of the simulated network leaving no time left for inference.
Similarly, when the SLA is 250ms this is just over the RTT of the network and the amount of time needed to run NasNet, our most complex model.
It should be noted that in our real Campus WiFi network CV is approximately 74\%.

\emph{Summary: \sysname is effective in handling highly variable mobile network by exploring a diverse set of deep learning models that expose different inference latency and accuracy trade-offs. When provided with very low SLA, \sysname should be used in conjunction with on-device inference and geographically-dispersed serving platforms to minimize SLA violations.}

\subsection{Decomposing the efficiency of \sysname}
\label{subsec:decomposing_benefits}

\begin{figure}[t]
    \centering
    \includegraphics[width=0.5\textwidth]{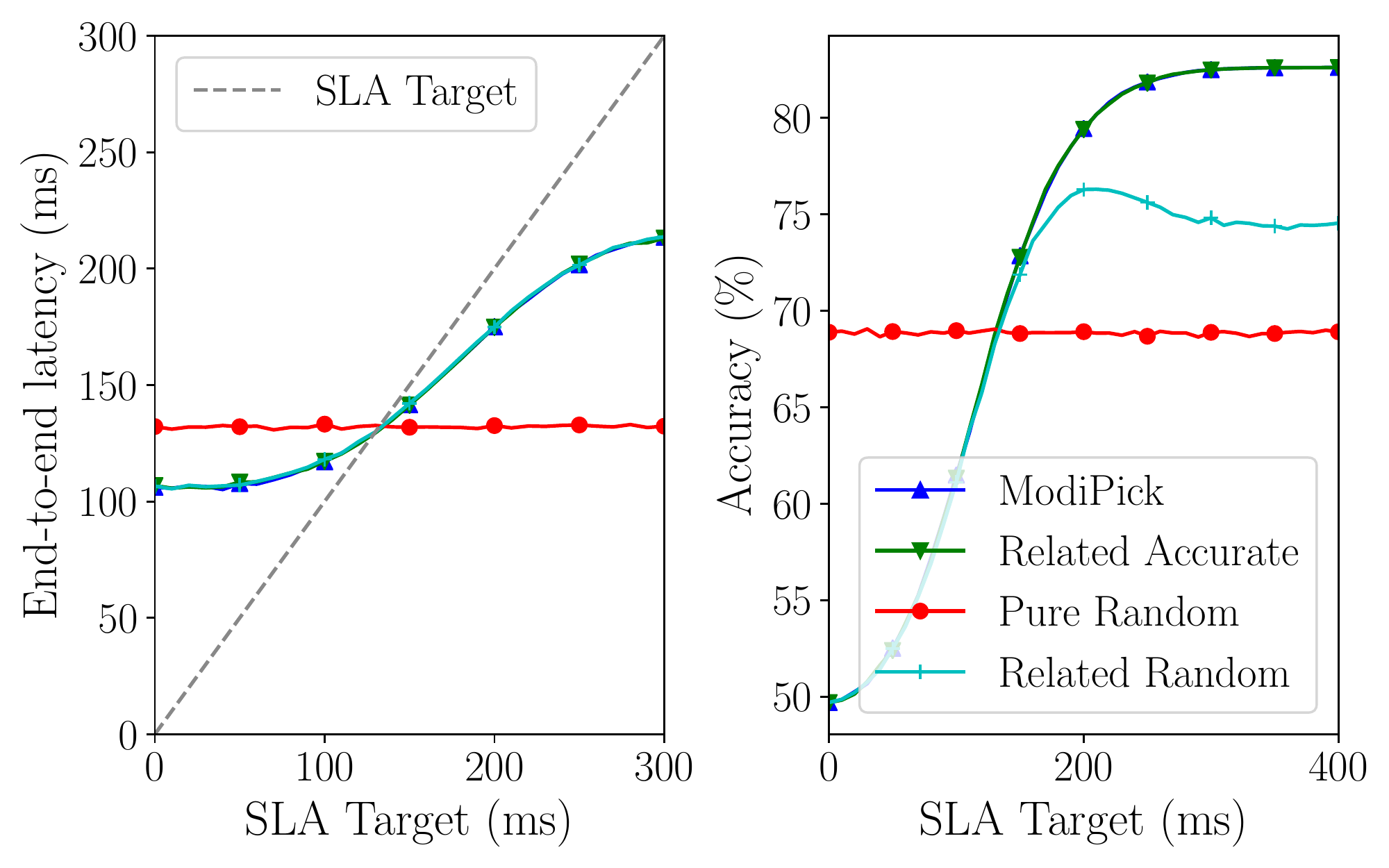}
        \caption{Decomposition of benefits of \sysname's three-stage algorithm. \textnormal{\sysname achieves similar accuracy and SLA attainment compared to \emph{related accurate}, indicating the effectiveness of our probabilistic approach. Both \emph{pure random} and \emph{related random} performs poorly in terms of inference accuracy due to their inability to distinguish models with different accuracy and latency, i.e., \texttt{NasNet Fictional} and \texttt{NasNet Large}.}}
    \label{eval:fig:algos:compare} 
\end{figure}

Last, we breakdown the performance benefits provided by \sysname by attributing to one of its stages of our dynamic accuracy-driven model selection algorithm (Section~\ref{subsec:modipickalgo}). We choose counterpart algorithms for each stage and evaluate the average end-to-end latency and inference accuracy achieved by all algorithms, in managing all twelve models listed in Table~\ref{eval:model-summary} We simulate a mobile network with average network latency of $50$ms and a standard deviation of $25$ms (CV = 50\%). 

For the first stage which selects the most accurate model constrained by its inference time distribution, we implement a \emph{pure random} algorithm that uniformly selects a model from all managed deep learning models. For the second and third stages where we probabilistically choose a model from the exploration set $M_E$, we use a \emph{related random} algorithm that uniformly chooses one model from $M_E$ and a \emph{related accurate} algorithm that picks the most accurate model from the exploration set. Note that we use a fictional model \texttt{NasNet Fictional} with the same inference time profile but with a low inference accuracy of 50\%, as a means to demonstrate the importance of stage two and three of our algorithm. 

Note, it is commonly assumed that the inference time and accuracy are positively correlated, i.e., the more accurate the model, the more time it takes to finish inference computation. However, in real deployment scenarios, model accuracy can fluctuate based on the actual inference requests served. In addition, as shown in Figure~\ref{fig:intro:cnnmodels}, deep learning models designed with different network architectures or at different time can also invalidate this assumption. Consequently, networks similar to \texttt{NasNet Fictional} can exist in deployment and it is important for \sysname to be able to explore them appropriately.

In Figure~\ref{eval:fig:algos:compare}, we plot the end-to-end latency (left) and the inference accuracy (right)  achieved by all four algorithms. As we can see, all three algorithms that choose from the exploration set $M_E$ are able to meet reasonable SLA target while \emph{pure random}  has approximately the same latency across all SLAs. This indicates that the construction of $M_E$, by stage one and two, is effective and enables good exploration opportunities to stay closely below SLA targets. The ability to adapt to increased SLA targets is important because it means we have the flexibility to use more accurate models. 

Similarly, as the SLA target increases, \emph{pure random} again achieves approximately the same inference accuracy across all SLAs. All three algorithms are able to gradually increase the inference accuracy by using slower but more accurate models from Table~\ref{eval:model-summary}. However, once we have a large enough SLA target ($\mathtt{\sim}150$ms), the exploration set $M_E$ is primarily consist of two models: \texttt{NasNet Large} and \texttt{NasNet Fictional}. At this point, \emph{related random} algorithm starts to experience inference accuracy decreases since it does not differentiate between these two models.  
Meanwhile, both \emph{related accurate} and \sysname are able to steadily improve inference accuracy by avoiding \texttt{NasNet Fictional}. 

Note there is only a negligible decrease in accuracy using \sysname when compared to \emph{related accurate} algorithm. This is because that \emph{related accurate} will always choose the most accurate model from $M_E$ while \sysname has a low probability of picking \texttt{NasNet Fictional} so as to update its model performance profile with controlled performance degradation. As mentioned before, models such as \texttt{NasNet Fictional} should not be completely ruled out from selection. The probabilistic behavior of \sysname is meant to allow for this exploration while generally maintaining accuracy, while \emph{related accurate} misses the opportunity to use models which may have improved accuracy or latency profiles.

\emph{Summary: \sysname's three-stage algorithm is effective in distinguishing and identifying the most appropriate model to use under dynamic inference conditions. All three stages contribute to and help \sysname combat the variable network conditions and improve inference accuracy safely.}

 \section{Related Work} 
\label{sec:related}

\sysname provides a cloud-based algorithm to improve the user-perceived inference performance for mobile deep inference requests. The key idea of \sysname is to mask the unpredictable network performance of mobile inference requests by leveraging a pool of deep learning models which can be used to trade-off latency and service quality measured as inference accuracy. 

Code offloading~\cite{maui, olie} has been widely used to augment the performance of mobile applications with constrained hardware resources. Due to the reliance on network connectivity, code offloading is often done at runtime~\cite{maui}. Determining the optimal partition of computation graphs can be solved optimally~\cite{maui} with approaches such as Integer Linear Programming (ILP). However, these optimal solutions fall short because they assume access to prior performance information, such as execution time and energy~\cite{maui, olie} and often incur long decision time. \sysname leverages the key idea of runtime computation offloading  with the unique aspect of selecting among various type of computations, i.e., different deep learning models, for both inference speed and accuracy gain. 

As deep learning models achieve unprecedented success~\cite{google_translate:one_shot, google_translate:ocr, nasnet} in classification tasks such as image recognition~\cite{resnet, DBLP:journals/corr/SzegedyVISW15, nasnet}, a lot of efforts~\cite{DBLP:journals/corr/HowardZCKWWAA17, nasnet} has been invested to integrate deep learning models to mobile application to provide novel features~\cite{google_translate:ocr, google_translate:one_shot}. Different from traditional machine learning based mobile applications, e.g., activity recognition~\cite{Ravi:2005:ARA:1620092.1620107} or energy prediction~\cite{6881749}, deep learning powered mobile applications raise additional challenges due to their heavy computation needs and resource limitation of mobile devices.

In particular, these models are often very ``deep'', e.g., can have more than 100s layers~\cite{resnet, DBLP:journals/corr/SzegedyVISW15}, and computation-intensive in nature. Directly running these models on resource-constrained mobile devices often result in very long inference execution time~\cite{tensorflow_models}, or even cause Out-of-memory errors~\cite{out-of-memory}. Mobile-specific model optimizations~\cite{DBLP:journals/corr/HowardZCKWWAA17, google_tensorflow:lite} that reduce computation needs or taking advantage of existing mobile hardwares, e.g., GPU or TPU,~\cite{pixel2.tpu} can shorten the inference execution time. However, such optimizations often involve sacrificing model accuracy~\cite{DBLP:journals/corr/HowardZCKWWAA17, tensorflow_models} or manual efforts to compile models to underlying hardwares that do not scale to heterogeneous mobile hardwares~\cite{google_tensorflow:lite}. 

Cloud-based solutions have demonstrated their effectiveness in handling heterogenous mobile capacity. In particular various model serving platforms~\cite{tensorflow_serving, clipper, DBLP:journals/corr/ChenLLLWWXXZZ15} provides web-based services that mobile applications can leverage. These platforms are often designed with the key focus of managing model lifecycle from training to deployment. \sysname complements such platforms with an intelligent model selection algorithm that gears toward mobile requests.

	To keep up with the increasing popularity of using deep learning within mobile applications, there has been a wide range of work on providing efficient mobile deep inference. These efforts range from optimizing mobile-specific models to improving the performance of inference serving systems.

\noindent \textbf{Mobile-specific model optimization.} As deep learning training aims at achieving human-level accuracy models are becoming increasingly complicated.  These complex models often have high computational demands and can consume significant energy. As a result, researchers have investigated various ways to make them more efficient~\cite{DBLP:journals/corr/SzeCYE17} via model optimizations. These efforts can largely be grouped into three categories.
First, post-training optimizations such as quantization uses simpler representations of weights and bins weights to improve compressibility~\cite{tensorRT, DBLP:journals/corr/HanMD15} allowing for reduced load time.
Second, techniques such as pruning~\cite{DBLP:journals/corr/HanMD15}, removing model weights with low contributions, reducing the number of computations needed for inference as well as model sizes.
Third, redesign of networks can also lead to improved inference time. An early example was the mobile-specific SqueezeNet~\cite{DBLP:journals/corr/IandolaMAHDK16} and this trend has continued with MobileNet~\cite{DBLP:journals/corr/HowardZCKWWAA17} which was designed as a compact alternative to the complex InceptionV3 model~\cite{DBLP:journals/corr/SzegedyVISW15}.

Although being effective at lowering inference latency, these optimizations often incur a degree of accuracy loss. For example, MobileNets can have over 20\% lower accuracy compared to InceptionV3. However, these model optimizations can be used to present a range of latency and accuracy points to allow for smooth trade-off between the two.

\noindent \textbf{On-device execution.}
Enabling deep learning models to be executed directly on mobile devices has been a goal of many researchers.  They have taken approaches such as using simplified model architectures~\cite{8454307, DBLP:journals/corr/IandolaMAHDK16, DBLP:journals/corr/HowardZCKWWAA17, DBLP:journals/corr/RastegariORF16} or reducing the complexity of existing networks~\cite{DBLP:journals/corr/HanMD15}.  These approaches generally sacrifice some accuracy in order to improve inference speed. In essence, these works target at either designing mobile-specific deep learning models or optimizing existing models to meet resource-constrained mobile platforms. In contrast, \sysname focuses on providing high-quality cloud-based inference service for mobile applications.

\noindent \textbf{Framework Redesigns.}
To enable running models across different hardware architectures, researchers have redesigned deep learning frameworks with the goal of providing optimized runtimes. 
For instance TensorFlowLite~\cite{google_tensorflow:lite} and Caffe2~\cite{caffe2} both leverage mobile-specific optimizations that allows deep learning models to execute smoothly on mobile hardwares. These optimizations, like those mentioned previously, are orthogonal to \sysname which improves mobile deep inference performance by judiciously selecting models at runtime.\par

\noindent \textbf{Remote Execution.}
There has lately been work regarding serving models.
It is not uncommon for inference to be done off of mobile devices either on closely located devices~\cite{Fang:2017:DEU:3131672.3131693} or on remote servers~\cite{tensorflow_serving}.
For remote servers, there are a number of platforms~\cite{tensorflow_serving, DBLP:journals/corr/ChenLLLWWXXZZ15} that aim to provide high-throughput serving for their models with low-level optimizations.  Alternatively serving systems such as Clipper~\cite{clipper} aim to provide high throughput for a number of different frameworks by transparently making the optimizations. These projects are beneficial to \sysname as they provide infrastructure supports for hosting a range of models. At the same time, \sysname complements these works by providing an automatic model selection that adapts to unpredictable mobile environments and frees application developers from the need to manually specifying model endpoints.

 \section{Conclusion}
\label{sec:conclusion}

The ability to provide efficient inference execution is crucial to enable a wide range of application scenarios that rely on increasingly complex deep learning models. Despite the recent advancement in on-device inference execution, cloud-based inference is here to stay for providing quality inference responses. In this paper, we design a dynamic cloud-based solution called \sysname that optimizes inference accuracy for mobile applications that rely on deep learning models. \sysname does so with a pool of models that expose different latency and accuracy trade-offs and by adapting its model selection to the heterogeneous mobile requirements given a predefined SLA, for each mobile inference request at runtime. In addition, \sysname leverages a probabilistic-based approach to explore the uncertain cloud inference execution time distribution and to combat the impacts of inaccurate inference performance estimation. Our evaluations show that \sysname is able to transparently switch between models with increasing SLAs and that \sysname achieves comparable accuracy while improving SLA attainment by 88.5\% as compared to greedy algorithm.

\section*{Acknowledgment}

This work is supported in part by National Science Foundation grants \#1755659 and \#1815619 and Google Cloud Platform Research credits.

\balance
\bibliographystyle{ACM-Reference-Format}
\bibliography{arxiv19_modipick}

\end{document}